\documentclass[doc]{apa6}

\usepackage{listings}
\usepackage{float}

\usepackage{amsmath,amssymb}
\usepackage{mathtools}

\usepackage{changepage}

\usepackage[utf8x]{inputenc}

\usepackage{textcomp,marvosym}

\usepackage{cite}

\usepackage{nameref,hyperref}

\usepackage[right]{lineno}

\usepackage{microtype}
\DisableLigatures[f]{encoding = *, family = * }

\usepackage[table]{xcolor}

\usepackage{array}

\newcolumntype{+}{!{\vrule width 2pt}}

\newcommand{\m}[1]{\mathbf{#1}}
  
\newcommand{\vs}[1]{\boldsymbol{#1}}

\title{Adaptive Synaptic Failure Enables Sampling from Posterior Predictive Distributions in the Brain}
\shorttitle{Adaptive Synaptic Failure}
\fourauthors{Kevin McKee}{Ian Crandell}{Rishidev Chaudhuri}{Randall O'Reilly}
\fouraffiliations{University of California, Davis}{Virginia Tech}{University of California, Davis}{University of California, Davis}

\abstract{
Bayesian interpretations of neural processing require that biological mechanisms represent and operate upon probability distributions in accordance with Bayes' theorem.
Many have speculated that synaptic failure constitutes a mechanism of variational, i.e., approximate, Bayesian inference in the brain.
Whereas models have previously used synaptic failure to sample over uncertainty in model parameters, we demonstrate that by adapting transmission probabilities to learned network weights, synaptic failure can sample not only over model uncertainty, but complete posterior predictive distributions as well.
Our results potentially explain the brain's ability to perform probabilistic searches and to approximate complex integrals.
These operations are involved in numerous calculations, including likelihood evaluation and state value estimation for complex planning.
}

\begin{document}
\maketitle

\section*{Introduction}
Bayesian interpretations of neural processing require that biological mechanisms represent and operate upon probability distributions in accordance with Bayes' theorem.
In this paper, we demonstrate how the random failure of synapses to transmit information may allow the brain to accurately represent multiple, compounding sources of uncertainty and perform accurate Bayesian inference.
%For a brief overview of Bayes' theorem, see the appendix.

%\paragraph{Bayesian neural computation}
%Bayesian theories of neural computation concern all of the above illustrated relationships between observations and inferences as they are implemented in the brain.
%We might say that the strength of the prior on $\theta$, i.e., its precision, influences the degree of change in $P(\theta|u)$ made by additional observations of $u$.
%In the brain, this is used to describe the effect of low-level sensory input on relevant high-level abstractions, and vice versa \cite{knill2004bayesian, friston2010free, friston2012history}.  
%It also provides benchmarks for efficient learning, as each new piece of information is weighed against what is already known \cite{aitchison2021synaptic}.

It has been shown that artificial neural networks can perform variational (i.e., approximate) Bayesian inference by randomly masking network weights, a form of \textit{dropout sampling}\cite{srivastava2014dropout, wan2013regularization, gal2016dropout,labach2019survey,gal2017concrete}.
Analogously, it is well established that synaptic vesicles randomly fail at a high rate to release neurotransmitters,\cite{allen1994evaluation,borst2010low,branco2009probability,huang1997estimating} leading to speculation that synaptic failure constitutes a mechanism of variational inference in the brain\cite{MonteroCosta19,MaassZador99,AitchisonLatham15,aitchison2021synaptic}.
In turn, some have demonstrated the plausibility of Bayesian neural computation by implementing generative neural architectures that emulate synaptic failure among other biological constraints \cite{guo2017hierarchical, neftci2016stochastic, mostafa2018learning}.
Whereas such models have previously focused on using dropout to sample over uncertainty in model parameters, we demonstrate that synaptic failure can also sample over posterior predictive distributions, of which parameter uncertainty is only one component.

%Related to the problem of approximate integration, this work also directly addresses the question of how the brain undertakes stochastic, global searches over the space of unknown variables, often compared to the simulated annealing algorithm for optimization \cite{KirkpatrickGelattVecchi83}.
%Random exploration of learned probability distributions, enhanced further by propagation of additional parameter uncertainty, may allow biological networks to escape from local attractor states that result from bidirectional feedback connections.

To understand the basic structure of a posterior predictive distribution, consider a model relating two observed variables $u_t, v_t \in \mathbb{R}$ at time $t$ that are jointly distributed $P(u_t,v_t)$.
The model takes a new observed input $u_{t+1}$ and uses parameters $\theta_t$, trained on all previous data up to time $t$, to generate a corresponding prediction $\hat v_{t+1}$ according to $P(\hat v_{t+1}| \theta_t, u_{t+1})$.
In Bayesian models, $\theta$ is randomly distributed conditional on finite vectors of previously observed inputs and outputs, $P(\theta_t| u_0\hdots u_t, v_0 \hdots v_t)$, i.e., model training is synonymous with inference of $\theta$ from past observations.
As such, $\theta$ is known imprecisely, with precision depending on its role in the model and the number of relevant observations up to the present, $t$.
As we are only interested in the distribution of the final prediction given some novel input, $P(\hat v_{t+1} |  u_{t+1})$, any model parameters $\theta$ are known as \textit{nuisance} variables.
To obtain $P(\hat v_{t+1} | u_{t+1})$, we marginalize out $\theta$, meaning we integrate over all of its possible values, each weighted by its respective likelihood.
If observations are independent and identically distributed, i.e., $P(u_t,u_{t-h})=P(u_t)P(u_{t-h})$ and $P(v_t,v_{t-h})=P(v_t)P(v_{t-h}), \forall h\neq t$, then the posterior predictive distribution is given as

\begin{subequations}
\begin{align}
    P(\hat v_{t+1} |  u_{t+1}) &= \int P(\theta_t,\hat v_{t+1} | u_{t+1}, u_0\hdots u_t, v_0 \hdots v_t)d\theta, \\
    &= \int P(\hat v_{t+1} | \theta_t, u_{t+1})P(\theta_t| u_0\hdots u_t, v_0 \hdots v_t)d\theta. \label{eq:ppd}
\end{align}
\end{subequations}

The total imprecision of a predicted output $\hat v_{t+1}$, defined by $P(\hat v_{t+1} |  u_{t+1} )$, thus includes internal sources of uncertainty, i.e. imprecision in $\theta$ defined by the parameter distribution, $P(\theta_t|u_0\hdots u_t, v_0 \hdots v_t)$ and external sources of uncertainty or the \textit{residual} distribution, $P(\hat v_{t+1} |  \theta_t,  u_{t+1})$, so named to denote random variation that remains after the outcome is conditioned on all available inputs.
For instance, if $P(\theta_t|u_0\hdots u_t, v_0 \hdots v_t)$ and $P(\hat v_{t+1} | \theta_t, u_{t+1})$ are both Gaussian, then the variance of the prediction is $\text{var}(\hat v_{t+1} | u_{t+1})=\text{var}(\hat v_{t+1} | \theta_t, u_{t+1}) + \text{var}(\theta_t | u_0\hdots u_t, v_0 \hdots v_t)$.

Approximate integration over posterior predictive distributions is likely to be particularly important to the field of reinforcement learning and models of human decision-making.
For instance, the method of Monte Carlo Tree Search (MCTS) involves estimating the value of a possible action by averaging over the expected returns from many simulated trajectories  \cite{SuttonBarto18}.
The expected returns are weighted by the likelihood of their respective states.
To assign accurate values, the agent must be able to average over not only the range of possible state predictions, but any uncertainty in the model parameters used to make those predictions.
The addition of parameter uncertainty would result in modulation of the breadth of the search based on prior knowledge, allowing a wider range of trajectories to be simulated where less is known in advance.
As a consequence of this work, agents designed to reflect neurobiology, such as spiking neural networks, may be able to plan and act according to both the complexity of the decision and the capacity of past experiences to inform it.

In this study, we aim to define a neural network constrained for biological plausibility that uses synaptic failure to draw random samples from $P(\hat v_{t+1}| u_{t+1})$.
We approach this by deriving separate dropout probability functions to sample from $P(\theta_t| u_0\hdots u_t, v_0 \hdots v_t)
$ and $P(\hat v_{t+1} | \theta_t, u_{t+1})$, then show that combining the two functions results in approximate samples from $P(\hat v_{t+1} | u_{t+1})$.
So far, dropout sampling of $P(\theta_t| u_0\hdots u_t, v_0 \hdots v_t)$ has only been formulated in biologically implausible contexts, such as with signed, Gaussian distributed weights.
We are tasked with deriving it for weights constrained between 0 and 1, representing generic bounds on synaptic efficacy.
It has not previously been shown in any context how synaptic failure may result in samples from $P(\hat v_{t+1} | \theta_t, u_{t+1})$, and hence, from $P(\hat v_{t+1}| u_{t+1})$ as a whole.

In the first section, we formulate an artificial neural network based on biological principles and subject to probabilistic interpretation that will be critical for our primary result.
Second, we find an analytic mapping from synaptic weights to transmission probabilities that allows representative samples to be drawn from $P(\hat v_{t+1} | u_{t+1})$, consistent with recent evidence that the rate of synaptic failures appears to be under adaptive control \cite{branco2009probability, borst2010low}.
Finally, we use simulations to demonstrate sampling from internal, parameter uncertainty, external, residual uncertainty, and from complete posterior predictive distributions in an abstracted network using only random synaptic failure.

\section*{Probabilistic neural network}
We present our model in five parts.
First, a biological, soft winner-take-all network model is outlined.
Second, we define a method of decoding neural activity in the network to obtain real posterior predictive samples $\hat v_s$ so that the network may be evaluated.
Third, we show how learning in the network is approximately inference of its weights from observed data.
Fourth, we use the learning principles to derive transmission probabilities to sample from the parameter distributions.
And finally, we derive a mapping from network weight values to transmission probabilities to sample from residual distributions.
By combining the fourth and fifth steps, we will obtain synaptic transmission probabilities that accurately sample the network's posterior predictive distribution.

Let us specify a neural network that senses the states of two real stimuli $u_t, v_t \in \mathbb{R}$ at time $t \in 1\hdots T$.
For each new input, $u_{t+1}$, the network generates a prediction, $\hat v_{t+1}$.
When presented with $u_{t+1}$, the network can represent uncertainty in $\hat v_{t+1}$ by drawing $S$ random samples from $P(\hat v | u)$.

The network consists of an equal number of input and output neurons $N$, respectively indexed $i, j\in 1 \hdots N$.
To represent $u, v$, input and output neurons fire at rates determined by their Gaussian receptive fields centered at locations $\vs \mu$ with equal widths $\sigma$.
Action potentials for each neuron are represented by Bernoulli random variables $x_i, y_j \in \{0,1\}$:

\begin{align}
P(x_i=1|u) &\coloneqq \text{exp}\left[ \frac{-(u-\mu_{i} )^2}{\sigma^2} \right], 
\quad P(y_j=1|v) \coloneqq \text{exp}\left[ \frac{-(v-\mu_{j} )^2}{\sigma^2} \right]. \label{eq:tuningcurves}
\end{align}

\begin{figure}[!t]
    \centering
    \includegraphics[width=\linewidth]{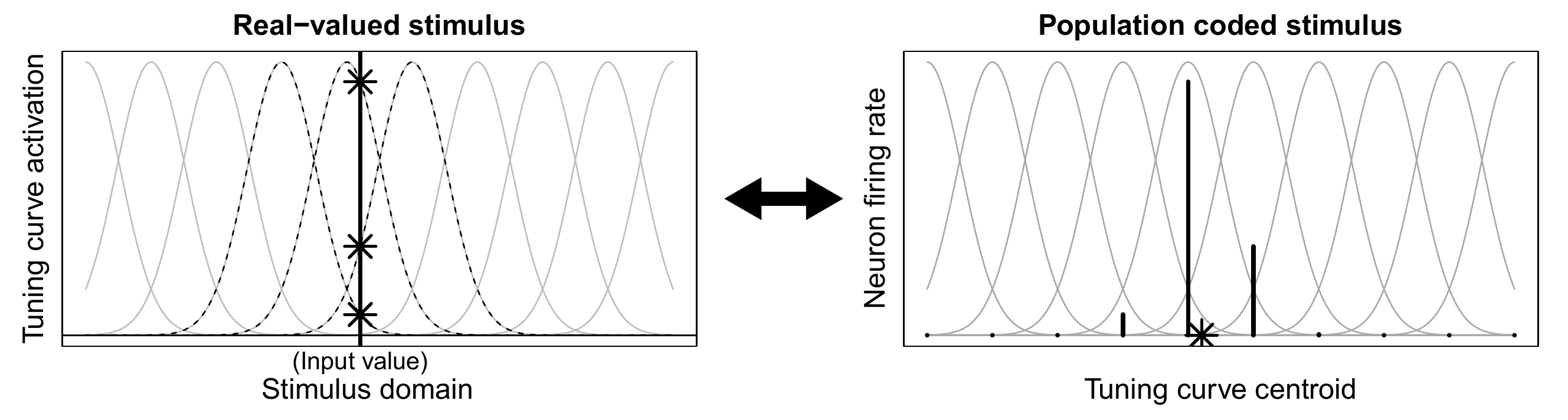}
    \caption{Dual representation of a stimulus as a real-valued number, the position of the vertical line on the left, and as neural firing rates, the heights of the lines on the right. Gray background curves represent the tuning curves of the sensory receptors. Asterisks shows how each representation is understood in the space of the other.}
    \label{fig:popcode1}
\end{figure}

%For each $u, v$ pair, a maximum of $R$ action potentials $x_{i,r} \in x_{i,1}, \hdots x_{i,R}$ and $y_{j,r} \in y_{j,1}, \hdots y_{j,R}$ are generated according to the above. 
The internal model is a linear, soft winner-take-all network connecting $x$ to $y$ by positive, bounded weights $w_{i,j} \in [0,1]$.
A posterior sample $s$ is represented by randomly masking weights with a Bernoulli random variable $m_{i,j,s}\in\{0,1\}$.
The degree of lateral inhibition active is given by parameter $\gamma$.
When $\gamma=1$, no lateral inhibition is active.
As $\gamma \to \infty$, $y$ approaches a one-hot vector or maximal sparsity in the output activations.
The model is defined as

\begin{align}
    P(y_{j,s}=1|\m W, \m x, \m M_s) &\coloneqq \frac{\left(\sum_i  m_{i,j,s}w_{i,j} x_{i}\right)^\gamma}{\sum_j\left(\sum_i m_{i,j,s} w_{i,j} x_{i}\right)^\gamma}. \label{eq:basemodel}
\end{align}

%For convenience, we reparameterize $\gamma$ in terms of $L\in [0,1)$ where $L=0$ represents no lateral inhibition:
%
%\begin{align}
%    \gamma &= \text{tan}\left(\frac{\pi(L+1)}{4}\right)
%\end{align}
%
%Hence, the full conditional distribution of the output is represented by,
%
%\begin{align}
%    P(y_{j,s}=1|\m w_{j}, \m x, \m m_{j,s}=1, L=0),
%\end{align}
%
%and a sample from the output distribution would be 
%
%\begin{align}
%    P(y_{j,s}=1|\m w_{j}, \m x, \m m_{j,s}, L>0),
%\end{align}

The neural network represents each stimulus pair $u, v$, and each sample prediction $\hat v_s$ by a series of action potentials.
The exact number of action potentials generated per stimulus value determines a third source of random variability in the output that is extraneous to this analysis.
Instead, we will use the expected values of $x_i$ and $y_j$ to represent standardized, asymptotic rate codes for both predictions and each posterior sample, with $\mathbb{E}[x_i | u] = P(x_i=1 | u)$, $\mathbb{E}[y_j | v] = P(y_j=1 | v)$, and

\begin{subequations}
\begin{align}
    \mathbb{E}[y_{j,s}| u] &\approx P(y_{j}=1|\m W, x_i=\mathbb{E}[x_i | u], \m M_{s}) \\
    &\approx\frac{\left(\sum_i  m_{i,j,s}w_{i,j} \mathbb{E}[x_i| u]\right)^\gamma}{\sum_j\left(\sum_i m_{i,j,s} w_{i,j} \mathbb{E}[x_i| u]\right)^\gamma}. \label{eq:basemodel}
\end{align}
\end{subequations}

The above gives the expected value of $y$ when $\gamma = 1$ and is a close approximation otherwise.
With this model, we leave weight dropout mask $m$ as the sole source of internal random variability.
We will show that, as a representation of synaptic failure, it is a sufficient mechanism to represent both Bayesian parameter uncertainty and residual uncertainty.

\begin{figure}[!ht]
    \centering
    \includegraphics[width=\linewidth]{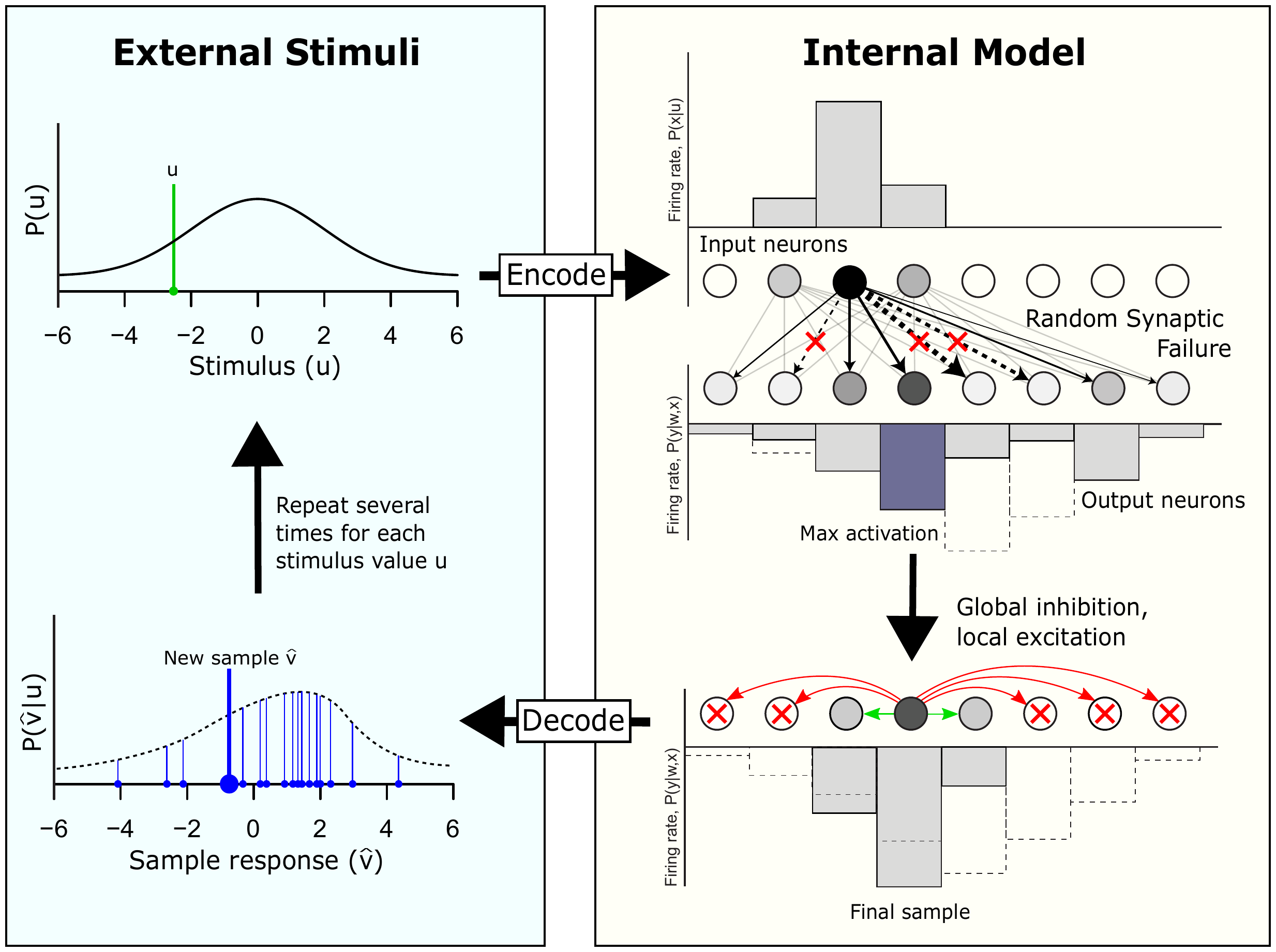}
    \caption{The neural network learns a distribution of responses or predictions $\hat v$ that follow stimuli $u$ by encoding them as neural activations $x$ and $y$. To sample from the distribution of $\hat v$, synapses ($w$) relating $x$ to $y$ randomly fail. Second, lateral inhibition results in selection of the most active neuron from the resulting subset.
    Third, local lateral excitation results in sustained activation of the nearest neighbors to the maximum, making $y$ a naturalistic population code for $\hat v$.
    We decode the samples by inferring $\hat v$ from $y$.
    By repeating this process, the whole distribution of $\hat v$ is represented over time.}
    \label{fig:samplediagram}
\end{figure}

Depending on the role of each layer in a multi-layer network, the few neurons that remain active after lateral inhibition may perform local lateral excitation.
This can be represented by Gaussian kernel smoothing of magnitude $\zeta$,

\begin{align}
    LE_j(\m z) &= \zeta\sum_k\text{exp}\left[ \frac{-(\mu_k-\mu_j)^2}{\sigma^2} \right] + (1-\zeta) z_j,
\end{align}

which can then be applied to our previously defined expected output as $\mathbb{E}'[y_{j,s} | u] =  LE_j(\mathbb{E}[\m y_s | u])$.
Lateral excitation of this form produces a naturalistic population code around the few remaining maxima among the output neurons.
If the receiving layer represents a previously observed stimulus that mediates the input and output layers, then in the absence of the mediating stimulus, the network may use lateral excitation to produce naturalistic samples in the mediating layer that have the same expected effect on the output layer as an observed mediator.
If the receiving layer is a hidden layer in a multilayer network, then lateral excitation induces spatial continuity among the learned representations.

\paragraph{Neural decoding scheme}
To test our hypothesis that the brain maps real input stimuli $u$ to implied real predictions $\hat v$ in accordance with Bayes' theorem, it is necessary to obtain sample predictions $\hat v$ by decoding each sample of neural activity.
In practice, decoding has no correlate in the biological theory \textit{per se}.
The brain only operates on its internal representations in terms of action potentials ($\m y$) and has no other way to represent continuous stimuli.

To infer each sample prediction $\hat v_s$ from the expected values of $x$ and $y$, we use grid search\footnote{Several local solutions are likely to exist in the divergence function for estimating $\hat v$. Optimization requires global search algorithms that are unlikely to converge sooner than a grid search in this case.} to choose $\hat v_s$ that minimizes the KL-Divergence of the output distribution from the probabilities given by the tuning curves:

\begin{subequations}
\begin{align}
  \hat v_s &= \text{arg min}\, \sum_j D_{\text{KL}}\left(P\left(y_j =1| \m W,  x_i = \mathbb{E}[x_i | u], \m M_s)||P(y_j=1|\hat v\right)\right), \\
%  &= \sum_j \text{D}_{KL}(m_j P(y_j|\m W, \m x)||P(y_j|u))  \\
  &= \text{arg min}\, \sum_j P(y_j =1| \m W, x_i = \mathbb{E}[x_i | u], \m M_{s}) \ln \left( \frac{P(y_j =1| \m W, x_i = \mathbb{E}[x_i | u], \m M_{s})}{P(y_j=1|\hat v)} \right). \label{eq:decodeKL}
\end{align}
\end{subequations}

For our purposes, this decoding method can be used with or without lateral inhibition and excitation with only minor differences in decoding because the maximally active neuron remains the same in any case.

\paragraph{Synaptic plasticity as approximate inference}
For a network to be exactly Bayesian, learning of network weights must be equivalent to inference of their distributions from previous data.
Inference of network weights by Bayes theorem is given as
%With subscript $i$ implied for all $x$, $j$ for all $y$, $i,j$ for all $w$, and sampling disabled during learning, we have

%\begin{align}
%    P(w_{i,j,t}|x_{i,0} \hdots x_{i,t}, y_{j,0} \hdots y_{t}  ) &=\frac{P(x_{0} \hdots x_{t}, y_{0} \hdots y_{t} %|w_{t})P(w_{t})}{P(x_{0} \hdots x_{t}, y_{0} \hdots y_{t} )}\\
%    &=\frac{P(y_{t}=1|w_{t}, x_{t})P(w_{t-1}|  x_{0} \hdots x_{t-1}, y_{0} \hdots y_{t-1} )}{P(y_{t})}.
%%    &=\frac{(\psi_i w_{t})^{x_t y_t}(1-\psi_i w_{t})^{x_t (1-y_t)} P(w_{t}|  x_{0} \hdots x_{t-1}, y_{0} \hdots y_{t-1} %)}{P(y_{t})}\\
%\end{align}

\begin{subequations}
\begin{align}
    P(\m W_{t+1} | \m x_0 \hdots \m x_{t+1}, \m y_0 \hdots \m y_{t+1} ) &=\frac{ P( \m x_0 \hdots \m x_{t+1}, \m y_0 \hdots \m y_{t+1} |\m W_{t})P(\m W_{t})}{P( \m x_0 \hdots \m x_{t+1}, \m y_0 \hdots \m y_{t+1} )}\\
    &=\frac{P( \m y_{t+1} |\m W_t, \m x_{t+1} )P(\m W_{t} | \m x_0 \hdots \m x_{t}, \m y_0 \hdots \m y_{t})}{P( \m x_{t+1}, \m y_{t+1} )}.
\end{align}
\end{subequations}

This equation defines a recursive relationship between the prior and the posterior of the weight distribution which manifests in iterative updates to the weights given each new set of input and output activations.
The distribution of $\m W$ that satisfies this equation under all conditions, for any given neural network, will be analytically intractable in general.
Most often, artificial neural networks have used variational inference to produce approximately Bayesian results \cite{blei2017variational}.
The distribution of weights is learned by optimizing an approximate, parametric form, which is usually a Gaussian process.
In biological neural networks, such optimization procedures are not biologically plausible.
Rather, learning likely involves a combination of local Hebbian rules and predictive coding.
Further, weight values represent synaptic efficacy and are thus positive and bounded.
For biologically plausible weight uncertainty, we instead choose a parametric distribution that both satisfies these constraints and approximately satisfies Bayes theorem.

The soft winner-take-all network we have defined can be framed as approximately a mixture of multinomial distributions, with each weight vector $\m w_i$ acting as a component probability mass function weighted by its associated input $x_i$.
For this reason, the Dirichlet distribution is a suitable choice for the distribution of the weight values because it is the conjugate of the multinomial, i.e., the distribution of the outcome probabilities.
For the Dirichlet to exactly model the network weights, three simplifying conditions must be true:
(1) Weight vectors $\m w_i$ per input $x_i$ must be normalized to sum to one to represent probabilities;
(2) Sampling and lateral inhibition must be disabled (all $m_{i,j,t,s}=1$ and $\gamma=1$) in the likelihood term $P(\m y_{t+1}| \m W_{t}, \m x_{t+1})$.
(3) Only one input neuron and one output neuron can be active at a time per stimulus pair ($\m x$ and $\m y$ are one-hot coded).
Then, the likelihood term reduces to $P( y_{j, t+1}|  w_{i,j,t}, x_{i,t+1})=w_{i,j,t}$, and Bayes theorem gives us the Dirichlet distribution:

\begin{subequations}
\begin{align}
        P( \m w_{i,t+1} &|x_{i,0}\hdots x_{i,t+1},  \m y_{0} \hdots \m y_{t+1}  )  \\
        & \propto P(\m y_{t+1} | \m w_{i,t}, x_{i, t+1})P(\m w_{i,t}| x_{i,0} \hdots  x_{i,t}, \m y_0 \hdots \m y_t) \\
        &= \prod_j w_{i,j,t}^{x_{i,t+1}y_{j,t+1}}\prod_j w_{i,j,t}^{a_{i,j,t}} \\
        &= \prod_j w_{i,j,t}^{a_{i,j,t+1}}, \\
        \text{where } a_{i,j,t} &= \sum_{\tau=0}^t x_{i,\tau} y_{j,\tau},\quad x_{i,\tau}y_{j,\tau} \in \{0, 1\}.
\end{align}
\end{subequations}

If the first simplifying condition is false, namely that weights are normalized, then the model is unaffected because the output activations are normalized and thus invariant to changes in the mean and scale of the weights.
The second simplification, that dropout and inhibition are disabled, is permissible because sampling and prediction are taken to be separate processes from learning and so the mechanisms of such do not factor into the learning of that distribution.

The third simplification that $\m x$ and $\m y$ are one-hot coded will almost always be false, but under the special case of sparsely coded inputs and outputs that we have described so far, the resulting error is negligible.
Specfically, Eq \ref{eq:tuningcurves} specifies the tuning curves of our model as overlapping Gaussian functions over $u, v$.
The overlap induces joint distributions $P(x_i, x_h|u)$ and $P(y_j, y_k|v)$ to the extent that $\sigma > |\mu_{i}-\mu_{h}|$ and $\sigma > |\mu_{j}-\mu_{k}|$.
For small $\sigma$, $P(\m x|u)$ and $P(\m y|v)$ are approximately one-hot vectors with a one representing the neuron that is most activated by each stimulus, e.g., $x_i = 1, x_h =0$, for all $h \neq i$. 
Otherwise, continuous stimuli induce neighboring input neurons to encode nearly the same output distribution.
The result is that error from this approximation mainly produces a kernel smoothing effect over the output distribution $P(\m y|\m W, \m x)$ with the tuning curve function (a Gaussian curve here) constituting the kernel.
For inputs with widely distributed codes that combine unique output combinations, this simplification is less tenable than the previous two.
Our scope is therefore limited to the case of sparse codings.

Let us denote $a_{i,*}=\sum_j a_{i,j}$ and imply subscript $t$ for all terms.
The means and variances of the Dirichlet distribution are

\begin{align}
    \mathbb{E}[w_{i,j}] &= \frac{a_{i,j}}{a_{i,*}},\quad
\text{var}(w_{i,j}) =\frac{a_{i,j} \left(a_{i,*} - a_{i,j}\right) }{ a_{i,*}^2 \left(a_{i,*} + 1\right)}.\label{eq:Dirichletmoments}
\end{align}

The expectation of $w_{i,j}$ is just the approximate frequency with which postsynaptic neuron $j$ out-competes the rest, and the variance of $w$ scales inverse to the non-normalized, cumulative activation values in $a$.
Positive increments to $\mathbb{E}[w_{i,j}]$ result from $x_{i,t} y_{j,t}=1$ and negative increments from $x_{i,t}y_{k,t}=1$ for all $k\neq j$.
Both the size of these increments and the variance of $w_t$ shrinks as the cumulative number of activations $\m a_i$ grows.

In summary, we have specified a linear soft winner-take-all network based on biological principles.
In this model, learning is approximately inference of the weights using a Dirichlet distribution.
This model serves our purposes in three ways: (1) It simplifies our task of abstracting from biological networks and preserving their domain constraints; (2) Learning is Hebbian rather than reliant on any global objective functions; (3) It provides analytic expectations for the means and variances of synaptic weights with their form of dependence upon past data given by Bayes' theorem.

Recalling Eq \ref{eq:ppd}, the model's output, excluding the dropout mask, is distributed as  $P( \m y_{t+1} | \m W_t,  \m x_{t+1})$, and its weights are distributed as $P(\m W_t| \m A_t)$.
The posterior predictive distribution is then

\begin{align}
P( \m y_{t+1} | \m x_{t+1} ) &= \int_0^1 P(\m y_{t+1} |\m W_t, \m x_{t+1})P(\m W_t| \m A_t) d \m W_t.
\end{align}

Generally speaking, Monte Carlo sampling is used to approximate integrals of the above form.
As we now have a distribution for $w_t$, we can approximate the model's internal posterior predictive distribution by sampling from that distribution:

\begin{align}
 \int_0^1 P(\m y_{t+1} | \m W_t, \m x_{t+1})P(\m W_t| \m A_t) d \m W_t &= \mathbb{E}_{\m W_t\sim P(\m W_t| \m A_t)}[P(\m y_{t+1} | \m W_t, \m x_{t+1})].
\end{align}

However, the brain cannot be said to have a mechanism for randomly setting synaptic efficacy according to an exact Dirichlet distribution.
Rather, we will next show that accurate but approximate samples from $P(\m W_t | \m A_t)$ can be drawn by random synaptic failure, also known as weight dropout.

\paragraph{Parameter uncertainty by dropout}
To further simplify notation, let us consider all $w$ and $a$ in reference to a common input $i$ at a common time $t$.
Let $\m{\hat w}$ be a vector of fixed weights from input $i$, and let weight and sample-specific dropout mask $m_j\sim~\text{Bernoulli}(\phi_{j})$.
To implement the Dirichlet distribution model by synaptic failure, we need to find dropout distribution $P( m_j \hat w_j|\m a) \approx P( w_j| \m a)$.
We do this by equating the means and variances, i.e.,  
$\mathbb{E}[m_j\hat w_j]=\mathbb{E}[w_j]$
and
$\mathbb{E}[(m_j\hat w_j - \mathbb{E}[m_j\hat w_j])^2]=\mathbb{E}[(w_j - \mathbb{E}[w_j])^2]$.
Equating the means and solving for $\hat w_j$ gives 

\begin{align}
    \hat w_j = \frac{\mathbb{E}[w_j]}{\mathbb{E}[m_j]} = \frac{\mathbb{E}[w_j]}{\phi_j}.
\end{align}
 
Using the above, we can then equate the variances and solve for transmission probability $\phi_j$.

%\begin{subequations}
%\begin{align}
%    \mathbb{E}[(m\hat w - \mathbb{E}[m\hat w])^2] &= \mathbb{E}[(w - \mathbb{E}[w])^2]  \\
%    \hat w^2 \phi_j(1-\phi_j)&= \frac{ab}{(a+b)^2(a+b+1) }  \\
%    \left(\frac{a}{\phi_j(a+b)}\right)^2 \phi_j(1-\phi_j) &= \frac{ab}{(a+b)^2(a+b+1) }, \\
%   \phi_j &= \frac{a^2+ab+a}{a^2+ab+a+b} \label{eq:Dirichlet2dropoutvar}
%\end{align}
%\end{subequations}

\begin{subequations}
\begin{align}
    \mathbb{E}[(m\hat w_j - \mathbb{E}[m_j\hat w_j])^2] &= \frac{a_{j} \left(a_* - a_{j}\right) }{ a_*^2 \left(a_* + 1\right)}  \\
    \left(\frac{a_j}{\phi_j a_*}\right)^2 \phi_j(1-\phi_j) &= \frac{a_{j} \left(a_* - a_{j}\right) }{ a_*^2 \left(a_* + 1\right)}, \\
   \phi_j &= \frac{a_j a_* + a_j}{a_j a_* + a_*} \label{eq:Dirichlet2dropoutvar}
 %     \phi_j &= \frac{a_j\left(\sum_j a_j + 1\right)}{\sum_j a_j\left(a_j  +1\right)} 
\end{align}
\end{subequations}

This equation maps the total amount of training data involved in learning $\hat w$ to its transmission probability $\phi$ and hence the weight variances $\hat w^2 \phi (1-\phi)$.
In summary, $w$ is Dirichlet-distributed in abstraction, but we show that the brain can use an approximate \textit{implementation} of that distribution by dropout.
By tuning synaptic transmission probability $\phi$ to values that result in the same variances expected from the Dirichlet distribution, the network produces approximately Bayesian results.
\footnote{
In principle, we could go on to draw sample masks in which dropout instances are correlated according to the covariance structure given by the Dirichlet distribution.
Correlated dropout may be simulated using a Gaussian copula or other methods to map from continuous covariances to binarized data.
%One need only compute the expected covariance of the Dirichlet from $a_i,j$ and $a_{i,*}$, generate multivariate normal data from the corresponding correlation matrix, and then to binarize the data values.
%The thresholds for binarizing are computed by inputting $\phi$ to the standard, inverse cumulative normal function.
We do not include this because, for the number of competing neurons considered here, the weight covariances shrink to negligible sizes.
}

\paragraph*{Residual uncertainty by dropout}
We have specified a way to draw samples from an approximate distribution of the network parameters using synaptic failure probability $\phi$.
To draw samples from a complete posterior predictive distribution, we need a second set of transmission probabilities by which the network draws samples from the residual distribution, $P(\hat v_{t+1}| \theta_t, u_{t+1})$.
Only sampling network parameters results in a distribution of the maximum \textit{a posteriori} of the prediction, or the most likely predictive value, whereas the residual distribution can be understood as the whole variety of outputs $v$ observed contingent on each input $u$.
Here, we derive synaptic failure probabilities $q$, that result in samples from the residual distribution.
We do this independently of the previous result ($\phi$) and assume that there is no additional dropout probability aside from that which we are deriving here.
Let us use the same simplifying conditions $x_{i}=1, x_{h}=0$, for all $h \neq i$, implying input subscript $i$ for all terms.
The problem can be stated as

\begin{align}
    \frac{w_j}{\sum \m w } &\neq P( m_{j,s} w_j  >  m_{k,s} w_{k}),\quad \forall k\neq j, \quad
    m_{j,s} \sim \text{Bernoulli}(q_{j}),\label{eq:problem} 
\end{align}

unless we find transmission probabilities $\m{q}$ that resolve this inequality.
Intuitively, if transmission probabilities are simply uniform or proportional to the synaptic weights, then the likelihood of sampling the tails of the distribution, encoded by the smallest weights, vanishes as the size of the network increases.
We conjecture that because $\m{w}$ encodes the distribution from which samples of $\m y$ must ultimately be drawn, then $\m w$ is sufficient to determine $\m q$.

To find a mapping of weights to synaptic transmission probabilities, consider the weights in descending order, where $j,k\in 1,...,n$, and $w_1 > w_j \geq w_k > w_n$.
Let $p_j \coloneqq {w_j}/{\sum \m w }$ and let $q_j$ be the transmission probability for weight $j$.
The probability of sampling from the largest of all weights is just that weight's value: $q_1=p_1$.
For each successive weight of rank $j>1$, the encoded probability must be equal to the probability that weight $j$ transmits and all larger weights fail:
	
\begin{align}
    p_j &= q_j\prod_{k=0}^{j-1}(1-q_k),
\end{align}

where $q_0=p_0=0$.
Solving recursively for $q$ in terms of $p$, we find 

\begin{subequations}
\begin{align}
    p_1 &= q_1, \\
    p_2 &= q_2\left(1-p_1\right), \\
    p_3 &= q_3\left(1-p_1\right)\left(1-q_2\right)\\ 
        &=q_3\left(1-p_1\right)\left(1-\frac{p_2}{1-p_1}\right)  \\
        &=q_3\left(1-p_1-p_2\right),\\
    p_4 &= q_4\left(1-p_1\right)\left(1-q_2\right)\left(1-q_3\right)  \\
        &=q_4\left(1-p_1\right)\left(1-\frac{p_2}{1-p_1}\right)\left(1-\frac{p_3}{1-p_1-p_2}\right)\\
        &=q_4\left(1-p_1-p_2-p_3\right), 
\end{align}
\end{subequations}

and so on for all $p_j$, giving us

\begin{align}
    p_j &=  q_j\prod_{k=0}^{j-1}\left(1-q_k\right) = q_j\left(1-\sum_{k=0}^{j-1}p_k\right).
\end{align}

\paragraph{Proof: Transmission probability from synaptic weight}
We show that  $p_j = q_j\prod_{k = 0}^{j - 1}(1 - q_k)$ implies $p_j = q_j (1 - \sum_{k = 0}^{j - 1}p_k)$ via induction. 
The base case is trivial: when $j = 1$ then $p_j = q_j \implies p_j = q_j$.
We assume the implication is true for $j = l$, giving us the identity 

\begin{align}
p_l = q_l  \prod_{k}^{l - 1}(1 - q_k) \implies p_l = q_l \left(1 - \sum_{k}^{l-1}p_k\right) 
\end{align}

and show that this implies the truth of the statement for $j = l + 1$.

\begin{subequations}
\begin{align}
	p_{l + 1} &= q_{l+1} \prod_{k}^{l} (1 - q_k) \\
	&= q_{l + 1}(1 - q_l) \prod_{k}^{l - 1}(1 - q_k) \\
	&= q_{l + 1}\left\lbrace \prod_{k}^{l - 1}(1 - q_k) - q_l  \prod_{k}^{l - 1}(1 - q_k)\right\rbrace \\
	& = q_{l + 1} \left\lbrace \prod_{k}^{l - 1}(1 - q_k) - p_l\right\rbrace \quad \text{ Def'n of $p_l$}\\
	&= q_{l + 1}\left\lbrace \frac{p_l}{q_l} - p_l \right\rbrace \quad \text{ Def'n of $p_l$} \\
	&= q_{l + 1} \left\lbrace1 - \sum_{k}^{l - 1}p_k - p_l\right\rbrace \quad \text{Induction step} \\
	&= q_{l + 1}\left(1 - \sum_{k}^{n}p_k\right) = p_{l + 1} 
\end{align}
\end{subequations}

Using this result, the transmission probability for weight $j$ is therefore

\begin{subequations}
\begin{align}
    q_j &= \frac{p_j}{1-\sum_{k=0}^{j-1}p_k}   \\
    &= \frac{w_j}{\left(\sum \m w\right)\left(1-\sum_{k=0}^{j-1}\frac{w_k}{\sum \m w}\right)}  \\
    &=\frac{w_j}{\sum \m w - \sum_{k=0}^{j-1}w_k } \\
    &= \frac{w_j}{\sum_{k=j}^n w_k}.  \label{eq:mapping}
\end{align}
\end{subequations}

For more than one input neuron transmitting to a shared output layer, we conjecture that no general solution exists such that transmission probabilities are constant across combinations of active inputs.
The above single-input mapping simplifies our derivation because the expected output activation $P(y_j=1|\m w_i, x_i=1, x_h=0, m_{i,j,s}=1)=p_{i,j}$. 
For multiple input neurons, normalization of a particular weight is dependent on $x_i$ and no longer constant.
It may be possible that separate solutions exist for every possible combination of input activations over $\mathbf{x}$, in which case synapses must adapt their transmission probabilities rapidly with changing input.

When multiple inputs are active, transmission probabilities must account for the possibility of being out-competed by synapses from other inputs.
The simplest approximation is to compute Eq \ref{eq:mapping} for each vector of weights, then divide all probabilities by the total expected input, $\sum \mathbb{E}[\m x|u]$.

%We sought to draw samples from $P(\hat v_{t+1} | \theta_t, u_{t+1})$.
%We do this by decoding samples $P(\m y_{t+1,s} | \m W_{t}, \m x_{t+1}, \m M_{t,s})$ to produce samples from $P(\hat v_{t+1} | %u_{t+1})$.
%Using the Eq \ref{eq:mapping}, we are now able to produce a matrix of probabilities that accomplishes this.
%Returning to unordered subscripts for input neuron $i$ and output neuron $j$, we have
%
%\begin{align}
%    q_{i,j} &= \frac{w_{i,j}}{\sum_{k} w_{i,k}},\forall k \text{ such that } w_{i,k} \leq w_{i,j}.
%\end{align}
%
%In brief, we are calculating and scaling Eq \ref{eq:mapping} independently for each column vector $\m{q}_j$ from each %corresponding column $\m{w}_j$.

\paragraph{Combined uncertainty}
As we initially showed in Eq \ref{eq:ppd}, total predictive uncertainty arises in both model parameters and in the model output itself holding the parameters constant.
In this case, prediction of a real output $\hat v$ from a real input $u$ is mediated by the brain's internal representations as neural activity, $\m x$ and $\m y$.
Dropout-based samples from the network's internal posterior predictive distribution, $P( \m y_{t+1} | \m x_{t+1})$, must incorporate variance from both the parameter distribution, $P(\m W_{t} | \m A_{t} )$, and the residual distribution, $P( \m y_{t+1} | \m W_{t}, \m x_{t+1})$.
The final samples can then be decoded to obtain samples from $P(\hat v_{t+1}| u_{t+1})$.
As we have now derived separate synaptic transmission probability functions that implement both component distributions, the two functions can be multiplied to produce a final dropout probability that implements sampling from $P(\m y_{t+1} | \m x_{t+1})$, and with subsequent decoding, $P(\hat v_{t+1} | u_{t+1})$.

\paragraph{Complete model summary}
Using our results, the equations for a linear winner-take-all network that draws samples by synaptic dropout from a learned posterior predictive distribution are

\begin{subequations}
\begin{align}
    P(x_{i,t+1}=1|u_{t+1}) &\coloneqq \text{exp}\left[ \frac{-(u_{t+1}-\mu_{i} )^2}{\sigma^2} \right], \\
    P( y_{j,{t+1},s}=1|\m {W}_{t}, \m x_{t+1}, \m M_{t,s} ) &\coloneqq \frac{\left(\sum_i  m_{i,j,t,s}  w_{i,j,t} x_{i,t+1}\right)^\gamma}{\sum_j\left(\sum_i m_{i,j,t,s}  w_{i,j,t} x_{i,t+1}\right)^\gamma},    \\
    w_{i,j,t} &= \frac{a_{i,j,t}}{a_{i,*,t}},  \text{ where } a_{i,*,t}=\sum_j a_{i,j,t}, \\
    a_{i,j,t}  &=  a_{i,j,0} + \lambda\sum^t_{\tau=1} x_{i,\tau}y_{j,\tau}, \\
    m_{i,j,t,s}  &\sim \text{Bernoulli}(\vs \phi_{i,j,t}  q_{i,j,t}),\\
    \phi_{i,j,t} &= \frac{a_{i,j,t} a_{i,*,t} + a_{i,j,t} }{a_{i,j,t} a_{i,*,t} + a_{i,*,t}}, \\
    q_{i,j,t}    &= \frac{w_{i,j,t}}{ \sum \mathbb{E}[\m x_{t+1}|u_{t+1}] \sum_{k} w_{i,k,t}}, \\ 
    &\forall k \text{ such that } w_{i,k,t} \leq w_{i,j,t}. \label{eq:finalmodel}
\end{align}
\end{subequations}

As stated previously, the random masking of weights by $m$ is the only internal source randomness in the network and produces accurate representations of both parameter uncertainty and residual uncertainty in the posterior samples.
Where appropriate, lateral excitation may be applied to the output layer to produce output distributions of the same form as those elicited by real stimuli.

If dropout probabilities $\phi$ or $q$ are set to zero and the output layer represents real stimuli $v_{t+1}$ according to

\begin{align}
    P(y_{j,t+1}=1|v_{t+1}) &\coloneqq \text{exp}\left[ \frac{-(v_{t+1}-\mu_{j} )^2}{\sigma^2} \right], 
\end{align}

then the weights update according to

\begin{align}
        w_{i,j,t+1} &= \frac{a_{i,j,t}+x_{i,t+1}y_{j,t+1} }{a_{i,*,t}+\sum_i \sum_k x_{i,t+1}y_{k,t+1}}.
\end{align}

%    \m w_{i,t} &\dot\sim \text{Dirichlet}(\m a_{i,t}), \\

\section*{Network Simulations}
To test that our theoretical network accurately samples from a learned posterior predictive distribution, we visually and numerically compare simulated data to samples drawn from the posterior distribution encoded in the learned network weights.
If the network is accurately representing the complete Bayesian posterior, then the network samples should be distributed according to the original data but with additional variance corresponding to parameter uncertainty, i.e., inflation in regions with fewer data points.

Monte Carlo simulations were used to estimate the accuracy of samples from $P(\hat v_{t+1}|\theta_t, u_{t+1} )$ and $P(\theta_t| u_0\hdots u_t, v_0 \hdots v_t)$ using weight dropout with probabilities $\phi$ and $q$.
Weight samples from the Dirichlet distribution were used to obtain a semi-analytic comparison for weight precision.

\paragraph{Data}
Two data-generating models were used.
The first involved mapping discrete input values to output distributions with increasing variance, i.e., heteroskedasticity, to test the accuracy of the dropout model when sampling from distributions with specific, known variances.
That is, data were generated from continuous normal distributions as $v\sim \mathcal N(0, 0.2+0.2(u+4)), u\in \{-4, -2, 0, 2, 4\}$.
The second model involved both heteroskedasticity and bimodality over a continuous range to examine the overall versatility of the model. 
For this model, data were generated from a mixture distribution:

\begin{subequations}
\begin{align}
v&\sim p(u)f_1(u) + [1-p(u)]f_2(u), \quad u\in [-4, 4]  \\ 
p(u)&=\text{logit}^{-1}(u/2),  \\ 
f_1(u)&= \mathcal{N}(-2, 0.2),  \\
f_2(u)&= \mathcal{N}(u/4, 0.2 + 0.0625(u+4)).   
\end{align}
\end{subequations}

In this second scenario, one distribution has a fixed mean and standard deviation, but with a declining density, while the other has a positive linear trend in the mean, variance, and density.
For both data models, 20,000 rows of data were generated to produce asymptotic results.

\paragraph{Model parameters}
For all simulations, data vectors $\m u$ and $\m v$ were transformed according to Eq \ref{eq:tuningcurves} to produce two $N\times J$ matrices of firing rates ($P(x|u)$ and $P(y|u)$) using a kernel of 81 Gaussian curves, equally spaced in $[-6, 6]$ with $\sigma=0.25$.
Expectations of $x$ and $y$ were used in place of binary action potentials to avoid additional, extraneous sampling error in estimates $\hat v$.
Posterior samples of $\m y$ were decoded to $\hat v$ by Eq. \ref{eq:decodeKL}.

Initial values, i.e., priors, for the weights were set as $a_0 \sim \mathcal U(0.025, 0.026)$.
Samples of $w$ from the Dirichlet distribution were used to provide reference for accurate sampling of $P(w_t| a_t, b_t)$ by dropout.
The learning rate $\lambda$ was set to 0.025 to produce larger weight variances for the purpose of demonstration.

1,000 samples per $v$ value were generated per input value $u$, and 70 repetitions of the simulation were run to produce distributions of each estimated variance component.
%The primary model of interest used the fixed matrix of weight means, $w_t=\mathbb{E}[w_t]$, and relied on dropout to sample both $P(y_{t+1}|\m w_{t}, \m x_{t+1})$ and $P(w_t | a_t, b_t)$
%For each posterior sample, this weight matrix was masked by $m_{i,j} = \phi_{i,j}q_{i,j}$.

\begin{figure}[!ht]
%\begin{adjustwidth}{-1.5in}{0in} % Comment out/remove adjustwidth environment if table fits in text column.
\centering
    \includegraphics[width=.9\linewidth]{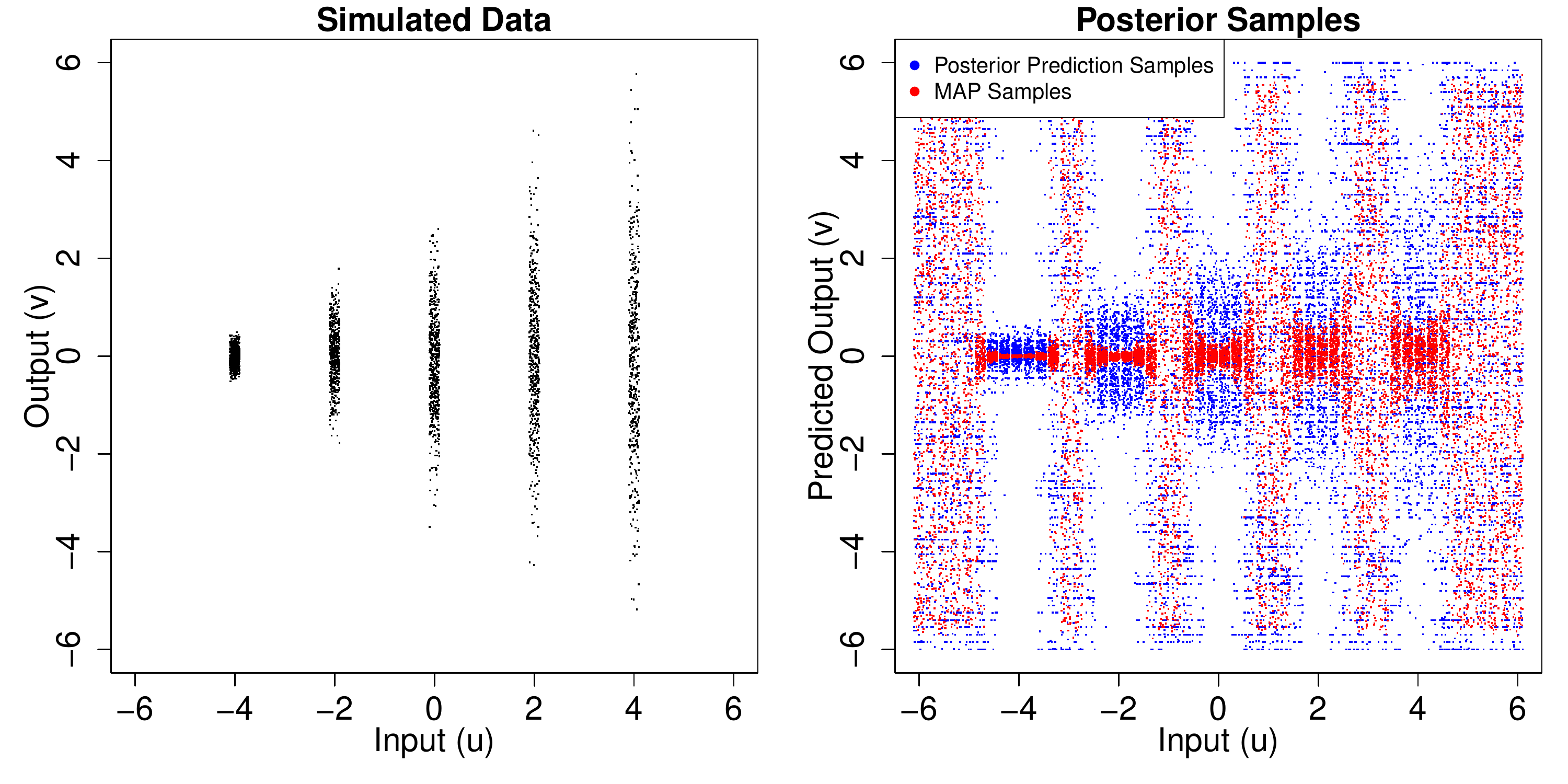}

    \includegraphics[width=.9\linewidth]{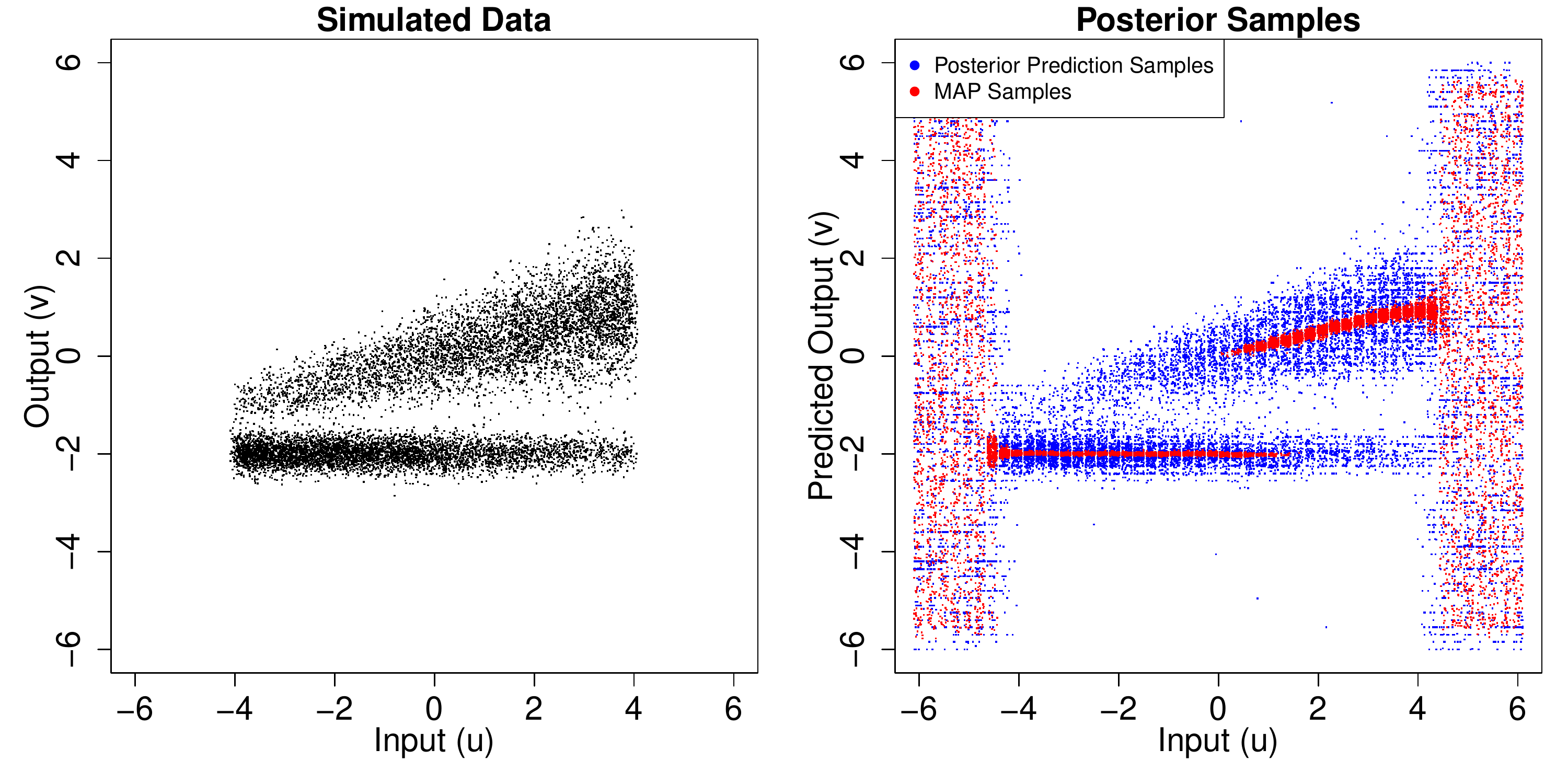}
    \caption{Simulated data versus dropout samples from the network. Top row: Changing variance model. Bottom: bimodal, heteroskedastic model.  Left column: Simulated data from each model. Right column: Samples from the complete posterior predictive distribution (blue) and samples of the MAP only demonstrating parameter uncertainty (red). X-axis jitter added to better show sample density.}
    \label{fig:simdata1}
\end{figure}

\paragraph{Results}
Fig \ref{fig:simdata1} shows the simulated data (left) with the associated posterior samples over the full domain of $x\in[-6,6]$ (right).
Samples from the complete posterior distribution are shown in blue.
The additional parameter uncertainty is overlaid in red as samples of the \textit{maximum a posteriori} (MAP).
The MAP was sampled by setting transmission probabilities to only $\phi$ such that the observed data distribution was excluded.

%\end{adjustwidth}

\begin{figure}[!ht]
    \includegraphics[width=\linewidth]{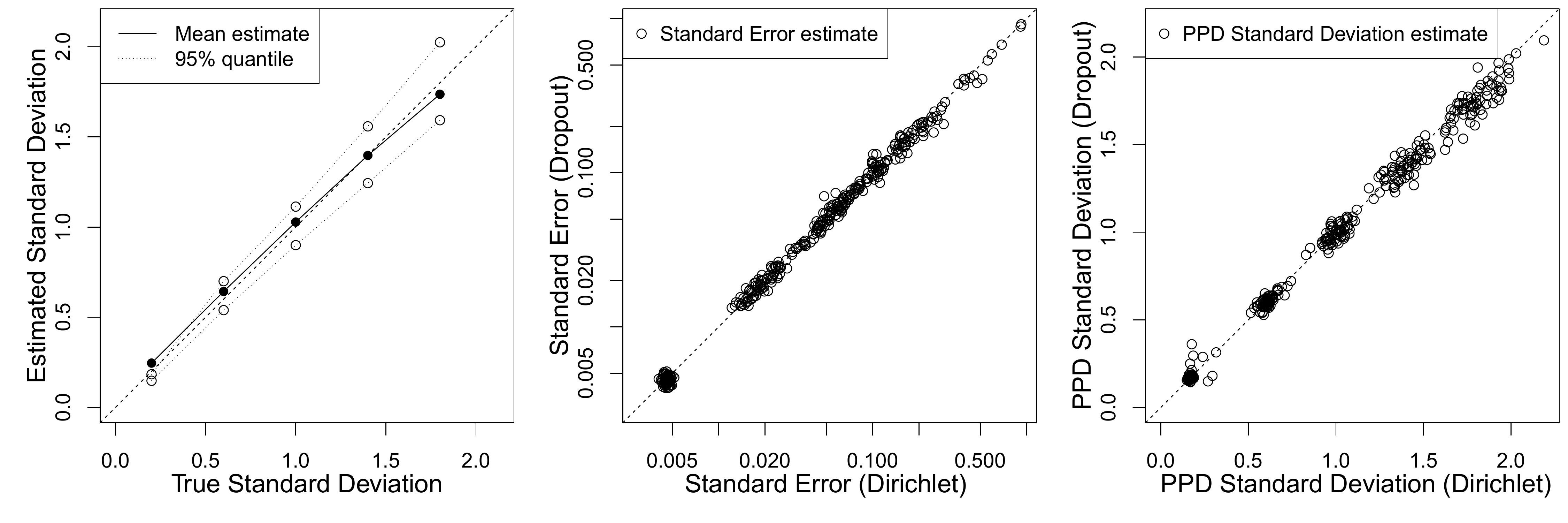}
    \caption{Left: Estimated standard deviations of samples from $P(y_j|\m w_j, \m x)$, excluding parameter uncertainty, versus the true, data-generating standard deviations.
    The standard deviations are nearly equivalent, showing that dropout with probability $q$ produces the correct residual distribution.
    Center: Parameter uncertainty comparisons as MAP standard error estimates obtained by sampling from the Dirichlet model (x-axis) versus by dropout (y-axis). The standard errors are also nearly equivalent, showing that dropout with probability $\phi$ produces the same final variance as drawing random weights from the Dirichlet.
    Right: Estimated standard deviations of full posterior predictive distribution $P(\hat v | u)$ from the Dirichlet model versus dropout, showing that the product $q\phi$ produces the same final posterior variance as sampling weights from the Dirichlet then applying dropout with probability $q$.}
    \label{fig:simresults}
\end{figure}

The network draws uniformly distributed samples where no data are available to inform input-output associations, such as between activated inputs values in the first simulation, and toward the edges of the input-output domains in both simulations.
The uniformity of the samples in these regions demonstrates maximal uncertainty in $w$.
Specifically, the network samples from the uniform priors of the network weights in the absence of any data to update those priors.
Conversely, the variance of MAP samples is smallest where the training data are most available, reflecting a high certainty in the most central or likely output value.
The gradations from maximal to minimal uncertainty visible around each input value in the first simulation reflects the weaker co-activation of tuning curves peripheral to each input value.

The first simulation was used to make numeric comparisons of the posterior distribution to the data-generating model.
The left panel of Fig \ref{fig:simresults} shows the average estimated residual standard deviations of output samples for each active input value.
These samples were generating using only transmission probability $q$, whereas parameter uncertainty was excluded.

The center pane of Fig \ref{fig:simresults} compares estimated standard errors, i.e., standard deviations of MAP samples, obtained by dropout with transmission probability $\phi$ to estimates given by samples from the Dirichlet distribution model.
Here, too, results fell along the identity line, showing that dropout produces a model that is accurate to the normative underlying Dirichlet distribution.

Finally, the right pane compares the estimated standard deviations of posterior samples between two models.
In the first (x-axis) weights were drawn from the Dirichlet distribution, followed by residual sampling via dropout with transmission probability $q$. 
In the second (y-axis), transmission probabilities $q\phi$ were used to sample from both the parameter and residual distributions in tandem.
This result shows that simply taking the product of the transmission probabilities for residual and parameter uncertainty is adequate to combine them and produce the full posterior predictive distribution.

\begin{figure}[!t]
    \centering
   \includegraphics[width=\linewidth]{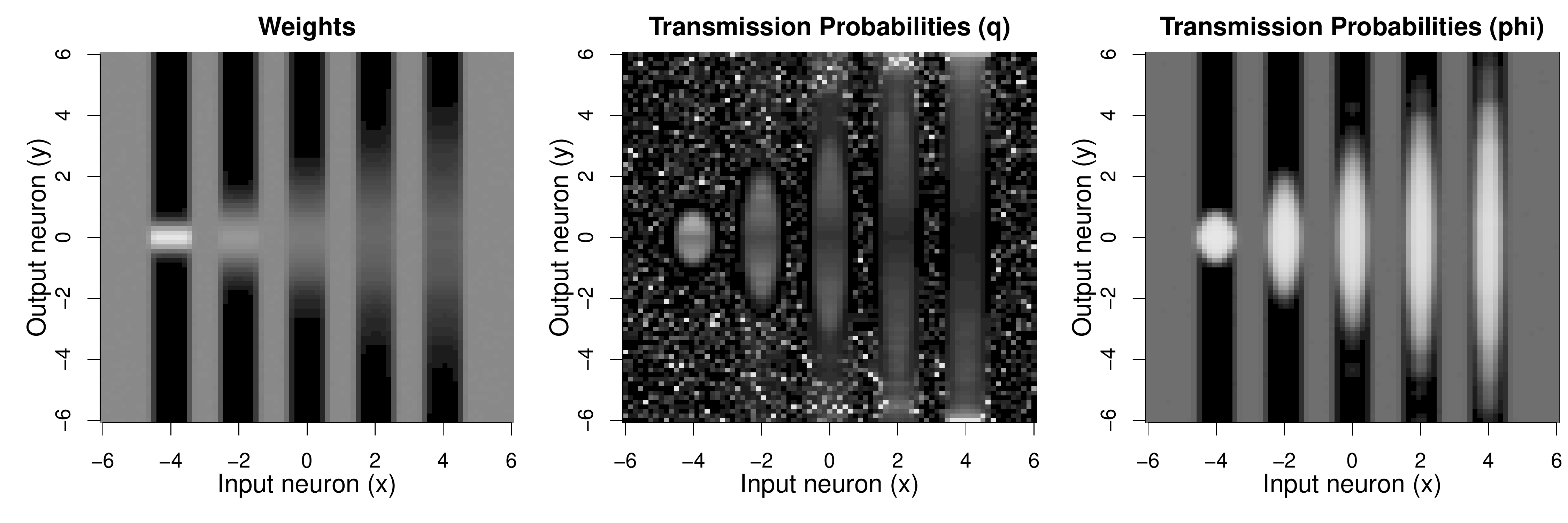}

   \includegraphics[width=\linewidth]{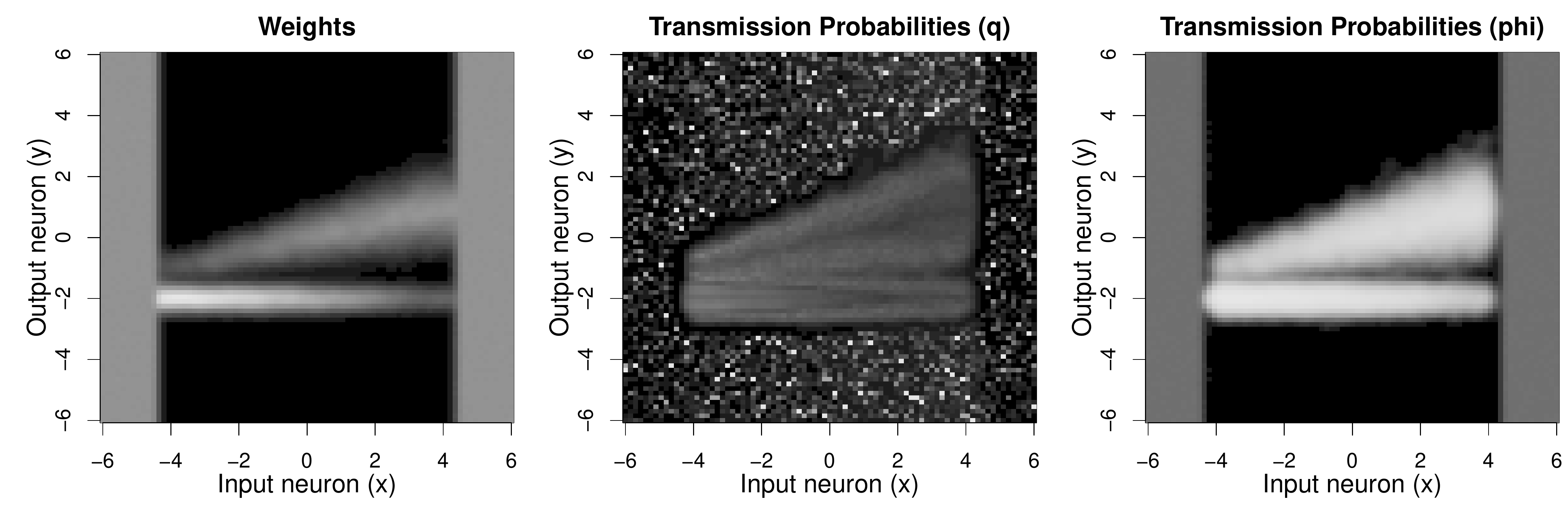}
    \caption{Left column: Network weights relating inputs  (x-axis) to outputs (y-axis); Center column: Transmission probabilities producing residual distributions; Right column: Transmission probabilities producing parameter distributions. Lighter corresponds to higher values in all plots. }
    \label{fig:weights}
\end{figure}

The matrices of learned weights and their release probabilities from a single iteration of simulation are shown in Fig \ref{fig:weights}, with lighter gray representing values closer to one in all plots.
In the first panels to the left, the data distributions are clearly encoded by the learned weight values for the active values of the input.
For input domains where no data were simulated, the priors appear as flat, uniform weights over the output domain.

In the center panels, transmission probabilities are shown to be highest among weights representing the tails of the distributions.
The heightened transmission probability in the distributional tails serves to counteract the inhibitory effects of the most active output neurons, which would otherwise prevent less likely network states from being sampled.
In coordinates where no data were observed, release probabilities only reflect the random noise of the priors.
%The smallest network weights show the same dropout tendencies here as were apparent in the previous simulations of local learning.
%Under the iterative, fixed exponent learning rule used here, the smallest weights are neglected in favor of those that most define the observed distribution.
Under the analytic mapping, the smallest weights transmit 100\% of the time, which results in the scattered white pixels shown, but this aspect is artifactual to our abstracted derivation and is not assigned any practical or theoretical importance.

In the right-hand panel, transmission probabilities $\phi$ for parameter uncertainty are shown. 
Probabilities higher or lower than the median value (0.5, gray) correspond to areas where more data were observed.
Unlike the weight matrices, precision can be seen to be taper down in the periphery of exactly where data were observed.
In other words, overlapping tuning curves result in shared statistical information among consecutive neurons and a degree of smoothing.

\section*{Discussion}
%%% SUMMARY:
Our simulations demonstrate that neural networks subject to biological constraints can represent and sample from complete posterior predictive distributions by synaptic failure alone when release probabilities are systematically related to synaptic weights.
To show this, we specified a linear winner-take-all network with neural activations and weights bound between zero and one.
These constraints both represent saturation points of synaptic efficacy and neural membrane excitation and lend to a probabilistic interpretation of the network as a hierarchy of mixture distributions.
Each network weight is defined as the conditional probability of an action potential being emitted by its receiving neuron.
We showed that under this probabilistic interpretation, uncertainty in weights can be modeled using the Dirichlet distribution and implemented with weight dropout.
Residual uncertainty, that is, random variation among outcomes associated with a common input value, were also sampled with high accuracy using transmission probabilities analytically derived under the same first-principles interpretation.

The analytic mapping from weights to release probabilities can be stated concisely: the release probability of each synapse is its strength normalized by the sum over itself and all weaker synapses. 
Intuitively, this means that the normalization factor is largest when there are many projections with just slightly weaker connections.
For a nearly uniform output distribution, the sorted transmission probabilities will follow a hyperbola.
This distribution is apparent as the sparse noise in the probability matrices shown in Fig \ref{fig:weights} where no data were available.
Further analysis may be warranted to rigorously generalize both our analytic derivation of release probabilities to the multi-input case, though we found that such rules derived from the single-input simplification appear to work well in simulation.
It is unclear how much tolerance to the inevitable but small approximation errors we should expect from the brain.

%Distinction: Normative or 'encoded' distribution versus implementation.
Our results rely upon a central distinction between abstracted, normative distributions that are derived via Bayes' theorem and dropout-based implementations that achieve an approximately Bayesian result.
That is, the brain probably does not have cellular mechanisms corresponding exactly to the operations of Bayes' theorem, but can produces results that happen to be well described by such.
Similarly, our approach shows how synaptic failure can be tuned to emulate the two most important statistics of the Dirichlet distribution, namely the means and variances of the network weights thus emulating the Bayesian operations from which it is derived.

%%% Biology
There is growing evidence that synaptic failure rates adapt purposely in the brain in support of computational theories such as ours.
Synapses in the mammalian neocortex have been found to exhibit high rates of communication failure, specifically the failure of presynaptic vesicles to release neurotransmitters \cite{allen1994evaluation}.
In vivo, release probabilities vary widely as a result of maturational differences, extracellular calcium, inhibition by ambient neurotransmitters, synapse type, and postsynaptic cell type, with most transmission probabilities falling well below 50\% \cite{borst2010low, branco2009probability}.
Huang \& Stevens \cite{huang1997estimating} estimated that two-thirds of synapses transmit subsequent to fewer than 17-33\% of presynaptic action potentials.
For our abstracted network to sample accurately, most transmission probabilities were below 20\%, though the overall statistics depends in practice on the number of neurons linked by lateral inhibition.

Synaptic concentrations of Ca\textsuperscript{2+} in particular have been shown to modulate vesicle release  \cite{dodge1967co}.
Evidence suggests that astrocytes respond to external stimuli and behavior with increased Ca\textsuperscript{2+} signaling by way of neurotransmitters such as noradrenaline, dopamine, and acetylecholine \cite{semyanov2020making, paukert2014norepinephrine}.
Hippocampal and cortical astrocytes modulate vesicle release probabilities and plasticity, and may be key to establishing biological control over statistical modes of processing.
The distal branches of astrocytes undergo rapid, externally induced Ca\textsuperscript{2+} transients as a function of their morphology, which changes in response to local neural activity \cite{semyanov2020making, bazargani2016astrocyte}.
Subsequent learning rules for synaptic failure probabilities should consider mathematical constraints based on the morphological and signaling dynamics of astrocytes at the synapses.
Likewise, our current findings may provide one functional interpretation of such activity.

In the models we have presented, parameter uncertainty is represented by mapping cumulative pre- and post-synaptic activity to an additional factor in the transmission probabilities.
Our mapping achieves the basic principle that as more evidence is gathered toward an inference, the precision of that inference improves.
However, by searching for the simplest mapping or learning rule for parameter uncertainty, we risk overlooking the possible advantages of situationally modulating parameter uncertainty.
Whereas observed distributions are shaped relatively slowly over the course of experience, parameter uncertainty may be modulated rapidly in response to external contexts and network states more broadly.
In dangerous circumstances, the network may increase transmission probabilities to limit sampling and act swiftly according to only the few most likely conclusions.

It has been suggested that Ca\textsuperscript{2+} signaling among astrocytes constitutes an additional, complementary pathway for longer-term information processing and modulation of mental states \cite{kastanenka2020roadmap}.
In theory, states of creative problem solving, idle thought, and rumination may be a few examples of posterior-sampling processes that are evoked by restricting Ca\textsuperscript{2+} and broadly reducing vesicle release.
In this mode, as we have shown, the network may conduct searches across its encoded distributions.
Stochastic search allows a network to jump out of one attractor state and explore others, much like the simulated annealing algorithm used in optimization \cite{KirkpatrickGelattVecchi83}.
By reducing probabilities below the optimal rates for sampling from observed distributions, a network can also sample from priors that extend beyond the boundaries of its encoded experiences.
This requires that such priors are true, literal priors, or preexisting synaptic connections that did not result from what we typically regard as learning.
In our results, this concept is visible in the form of samples drawn from a uniform distribution over the complete space of outputs where no input data were observed.

%Space vs Time
For simple models, sampling may be an unnecessarily slow mode of processing when the whole distribution is sufficiently encoded by the weights and hence by simultaneous neural activations in the output layer.
In more complex models, sampling may be necessary in the same way it is necessary for serial computing: to approximate a complex distribution over many unobserved variables. 
Population codes provide the capability to fully encode highly complex distributions, but in practice, applying those distributions to a particular action or perception may not be straightforward or possible without sampling.
Different intervals of the learned distribution over one layer may correspond to mutually exclusive actions.
Sampling allows the network to control its search across arbitrarily complex domains by engaging in random activity in the simpler, earlier stages of processing leading up to them.
This idea is analogous to sampling from the lowest-dimensional encoding layer of a variational autoencoder to search the final layer of complex reconstructions \cite{kingma2013auto}. 

%Reinforcement learning
Our findings may be useful for understanding any context in which the brain is required to integrate over complex, stochastic searches.
In neural circuits related to decision making and expected rewards, posterior sampling enables a form of Monte Carlo value estimation \cite{SuttonBarto18}.
In short, the value of a particular action or state is estimated by integrating over the values of all possible subsequent states and actions, each weighted by its likelihood.
In finite Markov decision processes, optimal choices can be analytically tractable, but in realistic settings, possibilities are often not clearly bounded.
Perception and inference must be used to judge the few most likely scenarios following each decision and integrate their value estimates accordingly.
Just as the posterior predictive distribution is defined by marginalizing over parameter uncertainty to make predictions, complex decision-making must involve marginalizing over the posterior predictive distributions of future states and actions to form accurate value judgments.
Development of reinforcement learning in spiking networks or other biologically plausible architectures may benefit the most directly and immediately from these results.

%% Limitations
% Hidden layers and variational inference
The models specified here differ from previous Bayesian brain models in a few ways.
First, this study did not involve hidden layers or the learning of of any generative models.
For many, generative variational models that learn to represent unobserved ``causes'' are the focal point of the Bayesian brain hypothesis \cite{knill2004bayesian, friston2010free}.
Such models originated with the Boltzmann machine\cite{hinton2007boltzmann}, Helmholtz machine\cite{dayan1995helmholtz}, and the variational autoencoder\cite{kingma2013auto}, and have been generalized in ideas like predictive coding\cite{RaoBallard99} and free energy \cite{friston2012history, friston2010free}.
Our probabilistic interpretation is not mutually exclusive with these model architectures, but remains to be synthesized with them.
By focusing on layers representative of two observed variables, we were able to derive and demonstrate Bayesian behavior under tractable first principles that allow for clear comparison with the true data-generating model.
In theory, the probabilistic relationships between input and output layer studied here should generalize to any two consecutive layers within a more complex, multi-layer network.

% Sampling is not involved in learning, separated here but they do not have to be.
Related to the problem of generative modeling, our model does not incorporate sampling into the learning process.
Generative variational models use sampling during training to regularize the learned latent spaces toward a simplified prior distribution \cite{kingma2013auto}.
As there were no latent spaces in this model, the weights could be learned directly from observed input-output pairings.
A reasonable next step for this theory is to generalize the abstract model to the case of one or more hidden layers.
The role of sampling and uncertainty in the learning process with regard to the learned structure of the hidden layers and relevance to variational inference can then be studied.

%Linear WTA vs Sigmoid nets
A third conceptual limitation may be the linear activation function of our model.
Whereas many network structures apply sigmoidal transformations to the integrated inputs of each receiving neuron, ours is simply linear combination of inputs followed by normalization to produce winner-take-all selection probabilities for each neuron.
It has been shown that winner-take-all networks with shallow architectures are capable of all the same nonlinear operations as sigmoidal networks \cite{maass2000computational}.
An ideal model of the brain might account for both sources of nonlinearity, as the sigmoidal function better emulates the thresholding behavior of action potentials. 
However, including sigmoidal transformations in the model introduces complexity into the relationship between layers that require our probabilistic re-framing of the network, and hence our Dirichlet-distribution based approximations, to be reconsidered.

% Time expenditure involved?
Another limitation of our model is that it implies lengthy computations that may take several seconds to complete. 
If the network requires posterior samples to be clearly distinguishable, then each posterior state (i.e., transmitted weight subset) must be sustained over hundreds or thousands of action potentials.
That is, it potentially takes more than a second to process each sample.
However, actual brain activity may instead be more fluid, allowing samples themselves to be uncertain as long as they, in aggregate, effectively convey the breadth of the encoded distribution.
Similarly, there is no reason to think that synapses fail in discrete, simultaneous sets as they do in our simulations, adding another element of overlap between consecutive posterior sample states.
In either case, it follows that the number of action potentials collected per sample likely constitutes a third component of uncertainty, namely signal fidelity, that is modulated primarily by the amount of time devoted to forming precise representations.

%% Future Directions
Questions of complexity and efficiency will be better answered with spiking networks programmed to adequately capture the analogue nature of brain states.
Future implementations of our model in spiking neural networks are planned, namely in the Axon framework that is currently under development \cite{axonGithub}.
Preliminary tests of an analogous network structure in the same data model presented here have been successful at recovering posterior predictive distributions from spike rates, but more work is needed to establish rules for adapting transmission probabilities with only locally available information according to biological findings and limitations.

\paragraph{Conclusion}
In summary, this paper demonstrates that synaptic failure can enable the brain to sample from Bayesian posterior predictive distributions.
The primary, overall implication of our result is that the brain can compute approximate integrals or perform stochastic searches over encoded posterior probability distributions. 
Approximate integration is necessary wherever a particular probability distribution in one neural population requires marginalizing over unknown values represented by additional, connected neural populations.
In this way, our findings supply theories of Bayesian computation in the brain with an elementary and broadly applicable operation.

\nolinenumbers

% Either type in your references using
% \begin{thebibliography}{}
% \bibitem{}
% Text
% \end{thebibliography}
%
% or
%
% Compile your BiBTeX database using our plos2015.bst
% style file and paste the contents of your .bbl file
% here. See http://journals.plos.org/plosone/s/latex for 
% step-by-step instructions.
% 
\bibliographystyle{apalike}
\bibliography{refs.bib}

\end{document}